\documentclass[twocolumn,showpacs,preprintnumbers,amsmath,amssymb]{revtex4}
\usepackage{graphicx}
\usepackage{dcolumn}
\usepackage{bm}

\begin{document}

\title{Comment on ``Strangelets as Cosmic Rays beyond the
Greisen-Zatsepin-Cuzmin Cutoff''}

\author{Shmuel Balberg}
\affiliation{Racah Institute of Physics, The Hebrew University, Jerusalem
91904, ISRAEL}

\pacs{98.70.Sa, 24.85.+p, 97.60.JD, 26.60.+c}

\date{\today}

\maketitle

In {\cite{MadLar03}} Madsen and Larsen propose that strangelets - stable
lumps of strange quark matter (SQM) - are candidates for the highest energy 
cosmic rays. They point out that the expected properties of strangelets make 
them easier to accelerate and less prone to energy losses than protons or 
ions. The purpose of this Comment is to point out that if ultra-high-energy 
cosmic rays (UHECRs) are indeed strangelets, their flux would guarantee that 
all neutron stars must be strange quark (SQ) stars, while no ``conventional'' 
(i.e., composed of nucleons and other baryons) neutron stars can coexist. 
This seems highly unlikely in view of various observed phenomena in neutron 
stars.

The hypothesis that SQM, composed of approximately equal abundances of up, 
down and strange quarks, is the true ground state of matter \cite{Bod71Wit84} 
has proved to be of special interest in the context of astrophysical compact 
objects. Self-bound SQM objects with masses of $M\lesssim 2\;M_\odot$ would be 
gravitationally stable and, for the most part, resemble conventional neutron 
stars \cite{AFO86}.  
While there have been attempts to identify observations of some neutron stars
as indication that they are SQ-stars, 
these were far from conclusive, and to date the existence of strange quark
stars remains speculative. 

If SQM is the ground state of matter, a strangelet  will 
absorb free neutrons, converting them to additional SQM (ordinary matter is 
repelled by positively charged strangelets, and so is unaffected). Hence, if a 
strangelet penetrates a neutron star, the object converts to a 
SQ-star on a time scale of minutes \cite{Olinto87}. Similarly, a massive 
stars which accumulates an SQM seed in its core will later give birth to a 
SQ-star, not a neutron star. The potential conversion 
of neutron stars into SQ-stars raises a key question: can both types of 
objects coexist in nature? If SQ-stars exist, then a
fraction of them must participate in binary mergers with other compact
objects, in which some SQM is injected into the galactic disk. The rate of 
binary meregrs in the galaxy appears to be large enough so that if SQM is 
sbsolutely stable, the flux of strangelets would 
ensure that all objects suspected as neutron stars are actually SQ-stars, 
Correspondingly, if some neutron stars are identified as being conventional, 
the inevitable conclusion is that SQM is not the ground state of matter and 
there are only conventional neutron stars \cite{Madsen88, CalFri91}. 

This argument might be undermined if the magnitude of SQM contamination
of the galaxy is suppressed, for example, if SQ-stars are disfavored in 
binaries (see \cite{POP02} and references therein). 
However, there can be no ambiguity in the context of a ``SQM as UHECRs''
scenario, for which the flux of strangelets is predetermined by 
construction. At $10^{20}$eV the flux is about one particle per 100 km$^2$
per year (see, e.g., \cite{UHECRs}). If UHECRs are strangelets, 
an existing neutron star should absorb several of them per year, and a massive 
star in the red giant stage would be impacted by as many as 
$10^{16}$ yr$^{-1}$. Moreover, if the strangelets are accelerated in 
astrophysical phenomena (e.g., the Fermi mechanism in shocks), the flux at 
lower energies must be orders of magnitude larger 
(the differential spectrum will be no harder than $dN/dE\propto E^{-2}$). 

In his reply \cite{reply}, Madsen argues that if a significant fraction of the 
strangelet energy is deposited in the strangelet as it impacts particles in 
the target star, high energy strangelets will disintegrate. In particular, if 
the binding energy per baryon is $E_B$, the upper limit for the minimal 
strangelet baryon number that will survive disintegration is $A_{min}=E/E_B$, 
where $E$ is the initial energy. The physics of strangelet fragmentation is 
essentially unknown, so the fraction of the initial energy which is deposited 
in the strangelet is unclear, as opposed to, e.g., pion production and 
acceleration of the target particles. A very high energy strangelet may also 
absorb baryons and grow in mass. But even if Madsen's upper limit is 
realistic, lower energy strangelets can be stopped prior to disintegration 
with relatively small baryon numbers. The key is that the total flux of 
strangelets (at all energies) required to have an ultra-high energy component 
at the observed level is so large, that even a minute probability for 
strangelet survival guarantees complete conversion of neutron stars to 
SQ-stars. Coexistence of both types of objects is ruled out. 

There do exist several observable indications that at 
least some neutron stars cannot be SQ-stars. In the more likely case of 
color-flavor-locked (CFL) \cite{CFL} SQM, the crusts of SQ-stars would 
certainly too thin to be consistent with the thick crust required to explain 
Type I X-ray bursts, in which light elements fuse to heavier ones in 
explosive flashes \cite{Xbursts}. Moreover, CFL SQ-stars probably cannot 
support millisecond rotation periods due to $r$-mode instabilities 
\cite{rmodes}. While non-paired SQM stars will have larger crusts that may be 
marginally consistent with these phenomena, they apparently
cannot glitch (undergo a sudden spin up observed in several pulsars; this is
the original argument of \cite{Madsen88,CalFri91}), and would also cool far 
too quickly to be compatible with observations \cite{Pageal00}. While the 
absence of a satisfactory model for SQ-stars which is consistent with all 
these observations cannot be considered as proof to the opposite, the more 
likely conclusion is that at least some neutron stars are, indeed, 
conventional. If so, the galaxy cannot be contaminated with strangelets at 
the level required for an UHECR-model.

I wish to thank K. Rajagopal for valuable discussions and advice.

\end{document}